\RequirePackage{amsmath}
\RequirePackage{amssymb}
\documentclass{llncs}[12pt]
\usepackage{llncsdoc}
\usepackage[all]{xy}           
\usepackage{fullpage}
\usepackage{hyperref}
\usepackage{eucal}
\usepackage{graphicx}
\usepackage{float}
\usepackage{epsfig}
\usepackage[usenames,dvipsnames]{color}
\usepackage{charter}
\usepackage{color}
\usepackage{comment}
\usepackage{fancyvrb} 
\usepackage{subcaption}

\begin{document}

\title{Socialbots whitewashing contested elections;\\ a case study from Honduras}
\titlerunning{Postelectoral whitewashing by socialbots}  
\author{E. Gallagher, P. Su\'arez-Serrato$^{1}$, \and E.I. Velazquez Richards$^{1}$}
\authorrunning{E. Gallagher, Su\'arez-Serrato, E.I. Velazquez Richards \and  M. Yazdani} 
%
%
\institute{$^{1}$Instituto de Matem\'aticas, Universidad Nacional Aut\'onoma de M\'exico, \\ Ciudad Universitaria, Coyoac\'an, 04510, M\'exico.}
\maketitle              
\begin{abstract}
We analyze socialbots active tweeting in relation to Juan Orlando Hern\'andez, the recently re-elected president of Honduras. 
We find a clear bimodal separation between humans and bots, using {\em Botometer} and its classifiers. 
Around one hundred separate communities of socialbots are identified and visualized, detected through the analysis of temporally coordinated retweets. 
\end{abstract}
\section{Introduction}
\noindent   On November 26th, 2017, a presidential election was held in Honduras. The central american nation of 9 million people has been in a state of turmoil since the election. Multiple protests have ensued calling for a new election to take place and for the previous results to be invalidated. Violent clashes ensued, claiming the lives of 31 people so far, and it remains a developing situation \footnote{See {\tt http://bit.ly/2lKpIcR}, {\tt http://fxn.ws/2DyUH3R}, \\ and {\tt http://bit.ly/2D8rLmx }.}.

\begin{figure}[ht!]
\centering
\includegraphics[width=\textwidth]{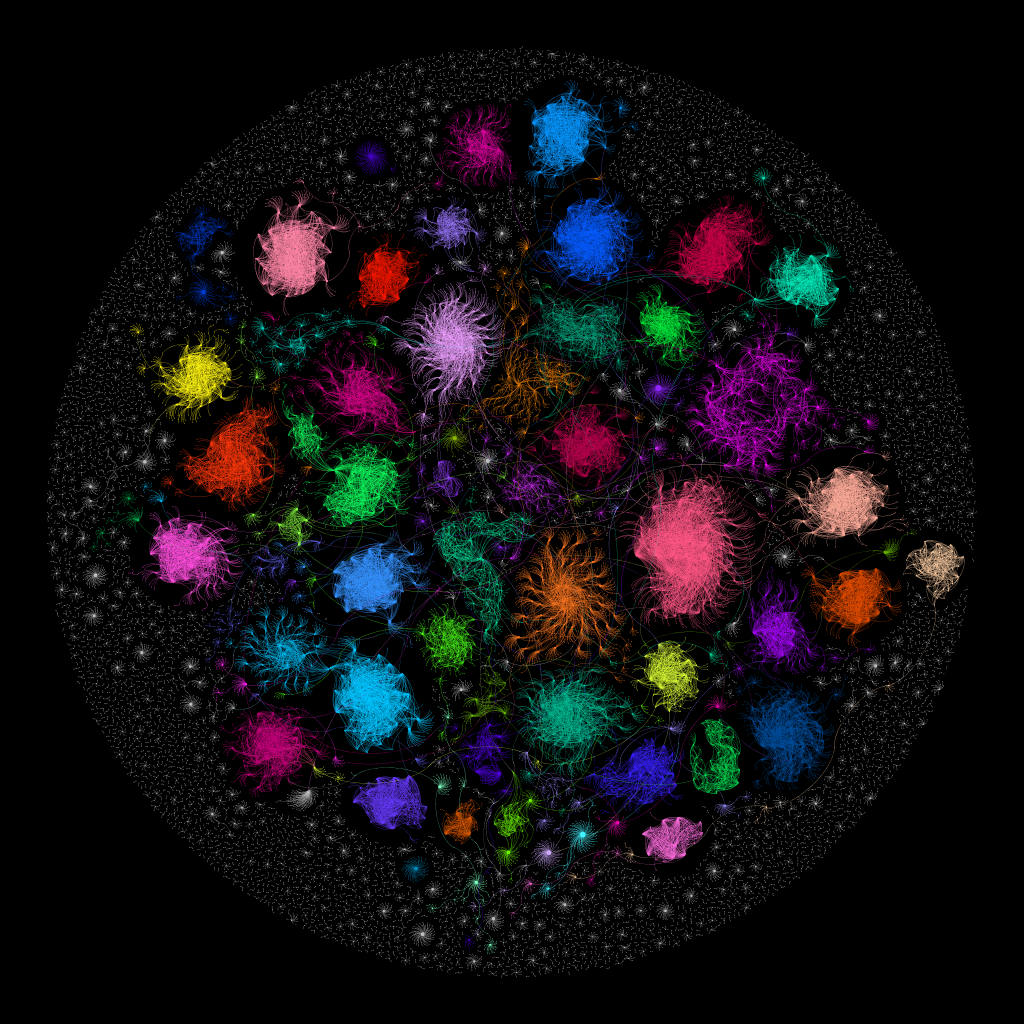}
\caption{ Full network of Twitter users mentioning the account during the collection period. Full Network: 41,288 tweets. Nodes: 26,363. Edges: 41,255. Communities: 4,108. \label{JOH-full}} 
\end{figure}

\begin{figure}[ht!]
\centering
\includegraphics[width=\textwidth]{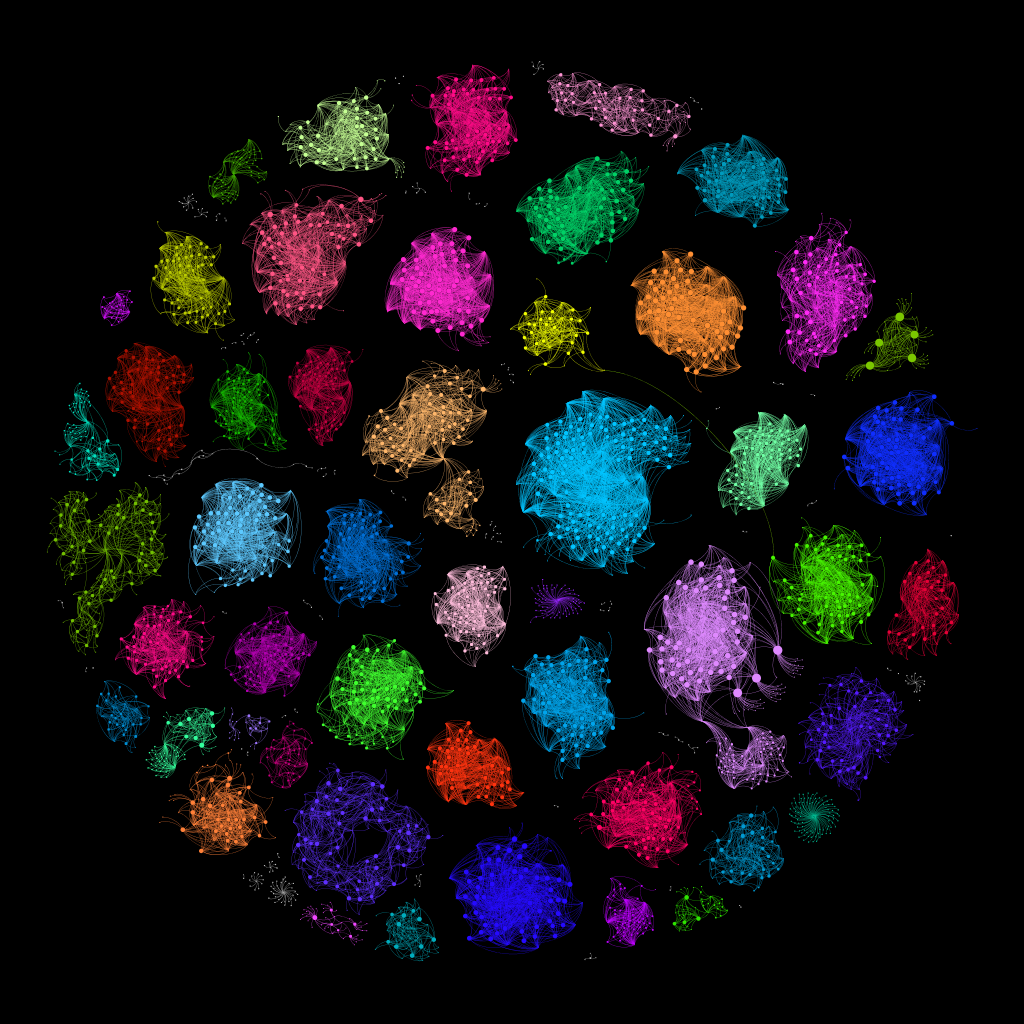}
\caption{ Subnetwork of Twitter users mentioning the account during the collection period whose accounts were managed using TweetDeck. TweetDeck Only: 22,519 tweets. Nodes: 3,767. Edges: 22,519.
Communities: 124. \label{JOH-tweetdeck}} 
\end{figure}

In this political context we obtained 41,288 tweets that mention the account of Juan Orlando Hern\'andez (@JuanOrlandoH). 
These were first analyzed visually using Gephi to understand how the users mentioning this account were related to each other. It can be seen in figure (\ref{JOH-full}) how the full network of mentions is distributed. Upon close examination of the dataset it was clear that a significant amount of these accounts were managed through TweetDeck (22,519 tweets from our dataset). 
In figure (\ref{JOH-tweetdeck}) the accounts that are being run through TweetDeck are visualized. More than half of the mentions of @JuanOrlandoH were generated by the accounts in this figure.
Furthermore, these two networks are obtained by using timestamps of tweets as nodes and accounts that retweet these as edges, so that the clusters form when there are coordinated, timed, retweets. 
This type of padding, or political astro-turfing, has been noticed and recorded before, but this is the first instance we know of where this scale of coordinated communities has been identified and visualized in a developing situation. 
A close reading of the accounts from figure (\ref{JOH-tweetdeck}) reveals that in some of the clusters the accounts even share a part of the username. 
The most notable in this sense are the Rivera and Santos teams. We also identified a group, which we called the Ladies team, of accounts with attractive women in their profile and whose main purpose is to emit sicophantic replies to @JuanOrlandoH.

\begin{figure}[ht!]
\begin{center}
\centering
\includegraphics[width=\textwidth]{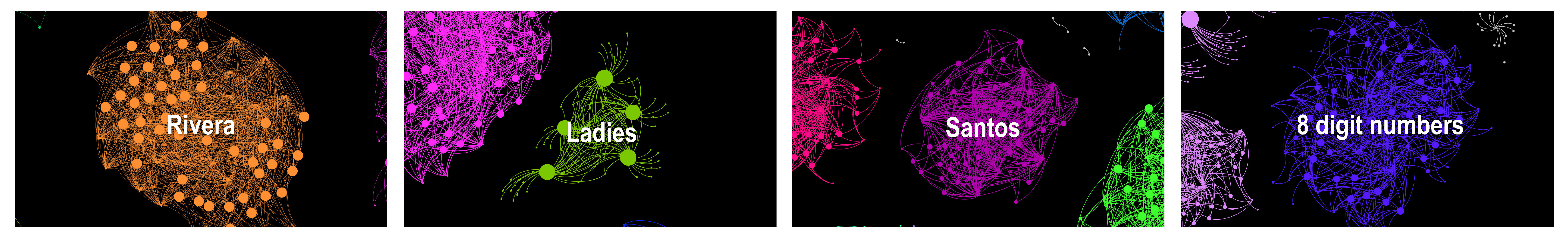}
\caption{ The socialbot accounts follow a similar logic in that they are created in batches within one to two days. Accounts in the Santos cluster were created either on December 6 or 8, 2017. Accounts in the Rivera cluster were created on either December 7 or 8, 2017. The accounts in the group we are calling Ladies were created on either June 10 or 17 of 2015. The socialbots follow each other and act in unison within their respective clusters. Each cluster retweets the same tweets at the same time, which is why we were able to easily graph their activity based on the timestamps of their tweets.\label{fig:teams}} 
\end{center}
\end{figure}

This rudimentary bot creation and management strategy makes these socialbots amenable to be identified. 
It is worrisome that the underlying message these socialbot accounts promote is that of good news, prosperity, and tranquility in Honduras. While in fact there are violent outbursts in the streets and wide-spread discontent with the way in which the last election's results are being imposed internally, and accepted internationally. 

\begin{figure}[ht!]
\begin{center}
\centering
\includegraphics[width=1\textwidth]{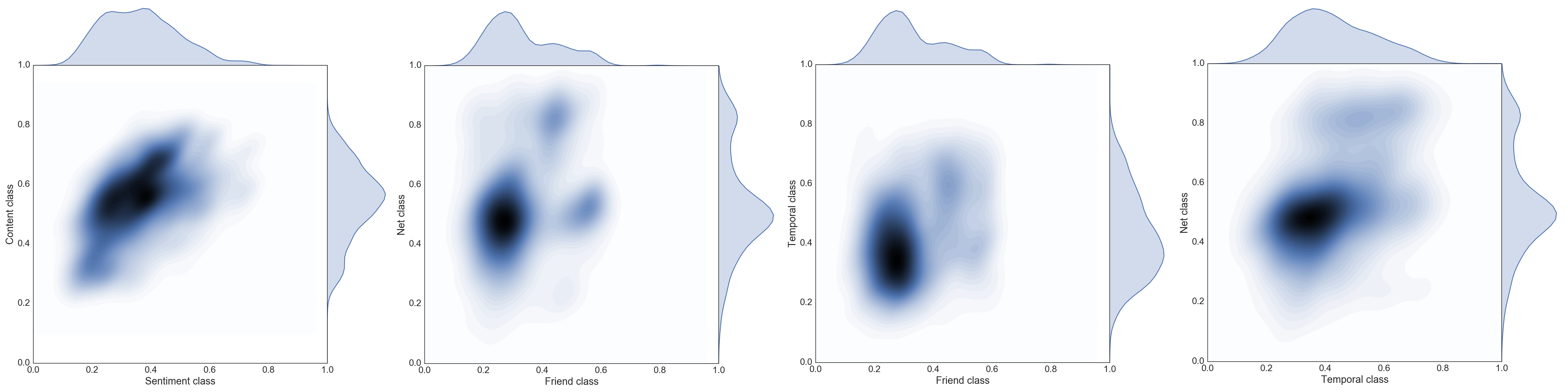}
\caption{ 2D Kernel decomposition estimate for Content-Sentiment, Network-Friend, Temporal-Friend, and Network-Temporal pairwise classifiers from Botometer, for Twitter users mentioning the account, for the sample obtained through Twitter's streaming API containing a total of 2,367 unique accounts.\label{fig:2d-kde}} 
\end{center}
\end{figure}

In order to validate the presence of socialbots we evaluated the accounts collected through {\em Botometer}. We find a considerable amount of bots acting and we identify a bimodal separation between human and bot users using Kernel Decomposition Estimate in 2 dimensions. Through the combination of non-language specific classifiers of {\em Botometer}, pairwise comparing its different classifiers classifiers, shown in figure (\ref{fig:2d-kde}).

\subsection{Background}
A review of {\em Botometer}, it's effectiveness and design can be found in \cite{RiseSocialBots2016}. The use of socialbots for political purposes and for spam can be found in \cite{Wooley2016}, \cite{KPM2013}. Other methods of bot detection have also been successful \cite{chavoshi2016identifying,Dickerson2014,Chu2012,Clark2015}. However {\em Botometer}, provides public API access which allows work like this to be carried out.

\section{Data preparation}

The total dataset comprises 41,288 tweets that mention the handle @JuanOrlandoH, collected via Tweet Archivist.  Dates/Times of capture ; Start: 12/25/2017 at 03:55:22 UTC; End:   01/01/2018 at 19:19:22 UTC. Out of the total dataset 22,519 tweets were sent using TweetDeck. The graphs from figures (\ref{JOH-full}) \& (\ref{JOH-tweetdeck}) are directed, layout algorithms OpenOrd \& Force Atlas 2 were used in Gephi for visualization of both these datasets. 

\section{Network Analysis}
We evaluated the centrality of the accounts being run with Tweetdeck, and found that the eigenverctor centrality created a proxy to naturally cluster different teams of accounts. We include in table (\ref{table:centrality}) the accounts with the highest eigenvector centrality, as it is possible that these accounts are controlling the rest, acting as botmasters and leaders whose activity is then mimicked by the rest of the bots.

\begin{center}
\begin{table}[htb]
\setlength{\tabcolsep}{8pt}
\begin{center}
\begin{tabular}{ll}
Andres\_Arguet  &   	1.0    \\
RicardoDuron15   &   	0.970588   \\
JeffreyChavez\_   &   	0.970588   \\
MelanyBulnes   &   	 0.941176  \\
AzulCerrato  &   	0.882353   \\
lauraso10566447   &   0.852941	\\
genesismargotc1   &   	0.852941   \\
\end{tabular}
\caption{Table of accounts with highest eigenvector centrality, indicating that they are possible botmasters\label{table:centrality}.}
\end{center}
\end{table}
\end{center}

\section{Conclusions}

We found almost a hundred clusters of coordinated socialbots acting to provide a positive social media fog in what turned out to be a violent post-electoral circumstance. While the socialbot accounts and the way in which they are managed is rudimentary, the implications for freedom of expression and control of dissenting voices in Honduras should be cause for concern. This study further highlights the need for social media companies to continuously monitor the abuse by automated accounts. Our findings underscore the need for tighter controls of social media abuse by regimes that seek to quell their opposition through a fake positive atmosphere online. 


\section*{Acknowledgments} We thank the OSoMe team in Indiana University for access to {\em Botometer }, and also Twitter for allowing access to data through their APIs. PSS acknowledges support from UNAM-DGAPA-PAPIIT-IN102716 and UC-MEXUS-CN-16-43.


\end{document}